\documentclass[]{aa}
\usepackage{natbib}
\usepackage{epsfig,lscape,rotating}
\usepackage{color}
\usepackage{amsmath}
\usepackage{amssymb}
\usepackage{ulem}
\usepackage{hyperref}

\renewcommand{\vec}[1]{\mbox{\boldmath$#1$}}

\begin{document}

\title{The magnetic diffusivities in 3D radiative chemo-hydrodynamic simulations of protostellar collapse}

\author{Natalia Dzyurkevich \inst{1}, Beno\^it Commer\c{c}on \inst{2}, Pierre Lesaffre \inst{1}, Dimitry Semenov \inst{3} }
  \institute{Laboratoire de Radioastronomie Millim\'etrique, UMR 8112 du CNRS, \'Ecole Normale
  Sup\'erieure et Observatoire de Paris, 24 rue de Lhomond, 75231, Paris Cedex
  05, France
\and
\'Ecole Normale Sup\'erieure de Lyon, CRAL, UMR 5574 du CNRS, Universit\'e de Lyon I, 46 all\'ee d'Italie, 69364, Lyon Cedex 07, France
\and
 Max-Planck-Institut f\"ur Astronomie, K\"onigstuhl 17, 69117
Heidelberg, Germany
      }


\authorrunning{N. Dzyurkevich et al.}
\titlerunning{Magnetic diffusivities in time-dependent chemo-dynamical collapse}

\date{}

\abstract
{
Both theory and observations of star-forming clouds require  
the simulations which combine the co-evolving chemistry, magneto-hydrodynamics and radiative transfer in protostellar collapse simulation.
A detailed knowledge of self-consistent chemical evolution for the main charge carriers (both gas species and dust grains) allows to correctly  estimate the rate and nature of magnetic dissipation in the collapsing core. Last is of crucial importance for answering the grand question of star and planet formation: the magnitude and spatial distribution of magnetic flux as the initial condition to protoplanetary disk evolution.}
{
We use a chemo-dynamical version of {\ttfamily RAMSES}, described in a companion publication, to follow the chemo-dynamical evolution of collapsing dense cores with the various dust properties and interpret the occuring differences in the magnetic diffusivity terms. Later are of crucial importance for the circumstellar disk formation. }
{We perform 3D chemo-dynamical simulations of 1 M$_\odot$ isolated dense core collapse for a range in the dust size assumptions. The number density of dust and it's mean size are affecting the efficiency of charge capturing and the formation of ices. The radiative hydrodynamics and dynamical evolution of chemical abundances are used to reconstruct the magnetic diffusivity terms for clouds with various magnetisation. 
}
{
The simulations are performed for a mean dust size ranging from $0.017\mu$m to 1$\mu$m, and we adopt both a fixed dust size and a dust size distribution. The chemical abundances for this range of dust sizes are produced by {\ttfamily RAMSES} and serve as an input to calculations of Ohmic, ambipolar and Hall diffusivity terms. 
Ohmic resistivity only play a role at the late stage of the collapse, in the innermost region of the cloud where gas density exceeds few times $10^{13}\rm cm^{-3}$. 
Ambipolar diffusion is a dominant magnetic diffusivity term in cases where mean dust size is a typical ISM value or larger.
We demonstrate that the assumption of a fixed 'dominant ion' mass can lead to one order of magnitude mismatch in the ambipolar diffusion magnitude.
 'Negative' Hall effect is dominant during the collapse in case of  small dust, i.e. for the mean dust size of 0.02 $\mu$m 
and smaller, the effect which we connect to the dominance of negatively charged grains. 
We find that the Hall effect reverses its sign for mean dust size of 0.1$\mu$m and smaller. The phenomenon of the sign reversal is strongly depending  on the number of negatively charged dust relative to the ions, and quality of coupling of the last to the magnetic fields.  
We have adopted different strength of magnetic fields,  $\beta=P_{\rm gas}/P_{\rm mag}=2,5,25 $. We observe that the variation on the field strength only shifts the Hall effect reversal along the radius of the collapsing cloud, but do not prevent it. 
}
{ The dust grain mean size appears to be the parameter with the strongest impact on the magnitude of the magnetic diffusivity,
dividing the collapsing clouds in Hall-dominated and ambipolar-dominated clouds, and affecting the size of the resulting disks. 
We propose to link the dust properties and the occurance and size of disk structures in Class 0 YSO's.  
The proper accounting for
dust grain growth in the radiative magneto-hydrodynamical collapse models shall be as important as coupling the dynamics of the collapse
with the chemistry.}

\keywords{chemistry --- hydrodynamics --- radiative transfer}


\maketitle

\section{Introduction}
The Ophiuchus, Serpent and Taurus-Orion clouds are rich in young stellar objects (YSOs), the total number is 
reaching towards five thousand. Nevertheless, the Class 0 YSO are rare among them due to the relatively short
 time window in which collapse of the cloud occurs. Additionally, those objects are quite distant and can be 
130 to 500 parsecs away. This and the fact that they are heavily veiled by the envelope makes the observations 
of their internal structure quite challenging. Nevertheless, there are at least four known Class 0 object 
possessing large ($>100$ AU) rotationally-supported disks: L 1527, VLA 1623A, RCrA IRS7B and HH212 MMS \citep{tobin15}.
More candidates has been added to this list in last year \citet{tobin16b}, whereas the difficulties are remaining
 to recover 
the rotation in the disks via molecular lines. The dense compact structures detected in the most of Class 0 objects
are small (less than 100 AU) and remain unresolved. The new powerful instruments allow to study the close-by objects 
in more details: the 
gravitational disk fragmentation resulting in triple binary in  L1448 IRS3B \citet{tobin16a}, 
the disk-envelope study made with ALMA for VLA 1623 \citep{murillo15}, 
the spiral arms in young disk in Elias 2-27 (also there are disagreements about its age) \citet{perez16}. 
Those observations indicate that there are more small disk cases versus more rare prominent large disks around YSO Class 0.

It is interesting to compare those trends to the results of the numerical simulations. 
Magnetic diffusivity terms are proposed as a way out of 
the magnetic braking catastrophe observed in ideal magneto-hydrodynamical (MHD) simulations of the collapse \citep[e.g.][]{henne08,com10}. 
Within the ideal MHD approach, the misalignment of rotation axis and magnetic field \citep{joos12}, 
the turbulence \citep{joos13,seifried15} and the reconnection processes \citep{lazarian12} can help to remove the magnetic flux.
In addition, the non-ideal MHD effects, including Ohmic diffusion, ambipolar diffusion and the Hall effect, are the natural consequence of gas condition in the dense molecular clouds and happen to reduce the magnetic braking efficiency, allowing for the  disk to form \citep[e.g.][]{li11,dap12,tom15,mas15,tsukamoto15b,tomida15,wur15}.

In the last twenty years,  a large number of
studies were performed 
 modelling the protostellar collapse and disk formation with non-ideal MHD, and
using a relatively simple fitting formulae for the ion and electron densities. 
\citet{krasno10} estimated the necessary Ohmic resistivity magnitude to be $10^{19}\rm cm^{2}s^{-1}$ 
in order to form a disk of radius 100~AU,
and showed also that Ohmic resistivity is likely much smaller. And indeed \cite{machida07,tomida15} show that Ohmic resistivity  becomes important deep in the cloud within the first hydro-static core (FHSC) when mean dust size is assumed rather tiny following the \citet{naka02}. 

Ambipolar diffusion 
is also shown to play an important role during the cloud collapse and disk formation. 
Its effect to redistribute the magnetic flux is strongest for low gas densities, where the magnetic field is coupled mostly to the ions and can ``slip'' through the neutral gas. 
The variety of studies ranges from theoretical and semi-analytical models with the 1D thin-disk approximation \citep{mestel56,basu04,krasnop02} to the 3D AMR simulations of the cloud collapse with the chemical look-up tables as in \citet{mas15,henne16}. 
\citet{duffin08,duffin09} considered an impact of ambipolar diffusion for dust-free gas in a one-fluid approach. Same as in \citet{mellonli09}, they found that the ambipolar diffusion does not weaken the magnetic breaking enough to allow the rotationally-supported disk to appear. Their finding were supported by \citet{li11}, who computed the ion density self-consistently from the Nakano chemical network \citep{naka02}. \citet{tomida15} performed 3D nested-grid RMHD simulations of the collapsing cloud and found that only tiny disk, less then 5 AU in radius, can be formed around a protostar during the first core collapse.
The simulations of \citet{mas15} and \citet{henne16} show that the formation of the Keplerian disk of 18 AU around 
the first core is a robust result of self-regulating magnetic fields under ambipolar diffusion possible in the 
collapsing cloud. 
In those models they include ambipolar diffusion as well as Ohmic dissipation, which are calculated with the method 
shown in \citet{mar16}. The similarities and disagreements between the latest simulations are discussed in \citet{mas15},
where disagreements seem to be caused by the different input parameters for chemical modelling of the gas. 

Thought it is less explored than ambipolar diffusion, the Hall effect is currently the subject of intense investigations \citep{li11,braiding12,tsukamoto15b,wur15}. 
According to \citet{li11}, Hall effect is dynamically significant but not capable of forming the rotationally-supported disk. The results of \citet{tsukamoto15} show the contrary, a possibility to form large disks under specific conditions.  
Its effect depends on the orientation between the rotation axis and the magnetic field (parallel or anti-parallel), which results in a bi-modality in the disk properties \citep{tsukamoto15b}. 
Altogether,  it is obvious that  magnetic diffusivities need to be taken into account all the way along the collapse,
to resolve the riddle of protostellar disks formation in the presence of the magnetic field. A detailed description of the dust-gas chemistry, the accurate treatment of processes  responsible for the ionisation, and the good model for the dust growth are thus highly desirable.

Chemistry is expected to play a crucial role during the star formation process. 
There are several reasons for making an effort of including the dynamical chemical 
evolution of key species into the radiative non-ideal MHD simulations of the molecular cloud collapse. 
Beside being a straight link to observations, the chemical abundances can be used to provide an 
additional cooling of the gas via atomic and molecular emission lines. 
Next very important aspect is that chemistry provides a realistic  degree of ionisation, 
ions distribution and the grain charging. As pointed out in \citet{libanerjee14,inutsuka12}, 
we need to consider all three types of magnetic diffusivity (Ohm, ambipolar and Hall) in order to 
resolve the long-standing problems of angular momentum and magnetic flux removal during the star formation.
Accurate calculation of the magnetic dissipations requires a precise description of the charge carriers, where chemistry must come in the calculations. 
While the cooling with molecular lines is expected to be important in the outer region of the envelope, the impact of chemistry on the magnetic diffusivities is going to be important everywhere, at any radius of the collapsing molecular cloud.

The equilibrium chemical models have been widely used to 
determine the degree of ionisation \citep[e.g.][]{naka02,mar16}. These authors 
deal mostly with representative ions (metals and non-metals), dust and electrons. Dust is 
important in the chemical models not only for forming complex molecules on its surface, but also
 for capturing electrons and neutralising through collisions with ions and charged dust, whereby 
the detailed physical properties of dust become very important \citep{mar16}. 
\citet{naka02}
assumed a dust size distribution dominated by tiny (a few nm) 
dust grains, which is in some contradiction with the observations of much larger grains in the dense cloud core, i.e. the so-called ``coreshine'' \citep{ste10}. 
The only known objects without observable coreshine, i.e. with no indications for large grains,  are in Gum/Vela region, 
where \citet{pag12} suspect a nearby supernova to be 
responsible for large dust destruction.


We have merged the \ttfamily{RAMSES }\rm code \citep{teys02} and the 
thermo-chemical Paris-Durham shock {\ttfamily PDS} code \citep{flower03,lesaffre05,lesaffre13,flower15}, with the technical description published in \citet{dzyurke16} (Paper I).
Paper I contains the detailed description of the chemical reactions as well as implementation details, which are also relevant for the simulations presented below. 
In this paper we ask the question about the coupling of the charged gas-dust mixture with magnetic fields, for various dust sizes.
We present the analysis of the magnetic diffusivity terms resulting from the dynamical chemical evolution for the radiation-(M)HD models of 
protostellar collapse. 
 There have been a few publications which can be considered as a precursor to our study. 
 \citet{tas12c,tas12a,tas12b} applied 
a complex chemical network to the isothermal phase of typical pre-stellar cloud 
collapse, performing a wide parameter study and searching for related pairs of 
neutral-ion molecules in order to provide sensitive probes of the importance of
magnetic fields and ambipolar diffusion in such clouds. We have considered the pairs of $\rm HCO^+,CO$ 
and $\rm H_3O^+, H_2O$ in the companion publication Paper I.
%
In spite of increased computational costs, the models such as presented in Paper I are  self-consistent and can be used threefold: for matching the observations, for calculating non-ideal MHD terms, and for the thermal chemical feedback (i.e. cooling with atomic or  molecular lines) \citep{hincelin16,harsono15,gerin15}. We  leave  the atomic/molecular lines cooling for the follow-up studies.
 The goal of this paper is to demonstrate the severity of the dust properties for the magnetic dissipation during the collapse. This is an important piece of information allowing to digest the previously performed collapse simulations with Hall and ambipolar diffusivities, and to make a theoretical prediction for the initial conditions for protoplanetary disks.

The paper is organised as follows.  
In section 2, we present the terse description of the reduced chemical network we use for protostellar collapse and the coupling of chemistry and radiation-hydrodynamics in  \ttfamily{RAMSES}\rm, whereas the details can be recovered in Paper I.  We present the setup for the collapse simulation and the analyse the magnetic diffusivities for fiducial model  in Sect. 3.  We discuss the effects of the various dust sizes on the  non-ideal MHD terms in Sect. 4. Section 5 concludes our work.


\section{Dynamical chemistry in {\ttfamily RAMSES}: the codes behind}

In this section we introduce shortly the chemo-dynamical modification of \ttfamily{RAMSES} \rm \citep{teys02},
 which we have described in detail in Paper I. 
We first describe the reduced chemical network which we use for protostellar 
applications. 
Then we describe the chemical equations in the context of \ttfamily{RAMSES}\rm.

\subsection{The reduced chemical network\label{sec:chemmod}}

\paragraph{Chemical species}
The chemical network we designed represents the main species and reactions necessary to describe CO abundances 
within mainly H-C-O chemistry in the early phases of protostellar collapse \citep{lesaffre05}.
We include 14 neutral species (H, H$_2$, He, C, CH$_{x}$ with $x=1,...,4$, O, O$_2$, H$_2$O, OH, CO, and CO$_2$), their corresponding single-charged positive ions, 
and ionised molecules CH$_5^+$, H$_3$O$^+$, HCO$^+$, H$_3^+$. In addition, iron is also included as a representative metal. 
For the details on the total elemental abundance or the dust elemental composition please consult the Paper I.
We constrain here to the description of the dust properties.

The size and mass of dust grains are calculated 
from the total mass of ``dust-core'' and ``dust-mantle'' elements and from the
adopted size distribution. 
The number of dust grains are calculated as $n_{\rm dust}=M_\mathrm{G,total}/ ((4/3)\pi \rho_{\rm solid}<a>^3$), where $\rho_{\rm solid}$ is the internal density, $M_\mathrm{G,total}$ is the total mass of the dust core elements, and $<a>$ is the mean radius calculated using  MRN size distribution $n(a) \propto a^{-3.5}$  \citep[][see appendix \ref{sec:ADust}]{mathis77}. Note that all dust core species  do not participate in the chemical reactions but are used exclusively to calculate dust grain mass. 

 All neutral species are allowed 
to freeze out on the dust grains, thus forming the grain mantles and increasing the weight of the grains. 
   Last, we consider three type of grains, namely neutral (G), single-charged positive (G$^+$), and single-charged negative (G$^-$). 
In total, the reduced chemical network describes the evolution of $N_\mathrm{species}=56$ species.

\paragraph{Chemical reactions \label{chemreact}}
We include gas-phase, freeze-out, and sublimation reactions. We do not take into account other types of reactions such as soft X-ray ionisation which is important for low density \citep[visual extinction A$_\mathrm{v}\sim0.2$,][]{wolf95}. For protostellar collapse applications, A$_\mathrm{v}$ is already larger than 10 for a typical mass of 1 M$_\odot$.

In total, our reduced gas-grain chemical network for H-C-O (and Fe as a representative metal) includes 231 reactions.

\subsection{The chemo-radiation-hydrodynamic model \label{crmhd}}

We have performed  3D time-dependent chemical calculations of protostellar collapse, using the time-dependent chemistry code {\ttfamily PDS} code  \citep{flower15,flower03,lesaffre05,lesaffre13} coupled with the 
{\ttfamily RAMSES}\rm\ code \citep{teys02}. For the technical details, we refer to the Paper I. 
Below we provide a brief summary of the code characteristics.

The set of chemo-RHD equations solved in \ttfamily{RAMSES}\rm \ with all radiative quantities estimated in the co-moving frame are
 \begin{equation}
\left\{
\begin{array}{llll}
\partial_t \rho + \nabla \cdot\left[\rho\vec{u} \right] & = & 0 \\
\partial_t \rho \vec{u} + \nabla \cdot \left[\rho \vec{u}\otimes \vec{u} + P \mathbb{I} \right]& =& - \lambda\nabla E_\mathrm{r} \\
\partial_t E_\mathrm{T} + \nabla \cdot \left[\vec{u}\left( E_\mathrm{T} + P_\mathrm{} \right)\right] &= & - \mathbb{P}_\mathrm{r}\nabla\cdot{}\vec{u}  - \lambda \vec{u} \nabla E_\mathrm{r} \\
 & & +  \nabla \cdot\left(\frac{c\lambda}{\rho \kappa_\mathrm{R}} \nabla E_\mathrm{r}\right) \\
\partial_t E_\mathrm{r} + \nabla \cdot \left[\vec{u}E_\mathrm{r}\right]
&=& 
- \mathbb{P}_\mathrm{r}\nabla\cdot{}\vec{u}  +  \nabla \cdot\left(\frac{c\lambda}{\rho \kappa_\mathrm{R}} \nabla E_\mathrm{r}\right) \\
 & &  + \kappa_\mathrm{P}\rho c(a_\mathrm{R}T^4 - E_\mathrm{r})\\
 \partial_t{n_x}/+\nabla \cdot [n_x \vec{u}] & = & - n_x \sum_r k_{rx}n_r  \\
  & & +  \sum_p \sum_q k_{pq}n_pn_q,  \label{chem_solver}
\end{array}
\right.
\end{equation}
\noindent where $\rho$ is the material density, $\vec{u}$ the velocity, $P$ the thermal pressure, $\kappa_
\mathrm{R}$ the Rosseland mean opacity,  $\lambda$ the radiative flux limiter \citep[e.g.][]{min78}, $E_\mathrm{T}$ the total energy $E_\mathrm{T}=\rho\epsilon +1/2\rho u^2 + E_\mathrm{r}$ ($\epsilon$ is the internal specific energy), $\kappa_\mathrm{P}$ the Planck opacity, $E_\mathrm{r}$ the radiative energy, and $\mathbb{P}_\mathrm{r}$ the radiation pressure. 

Note that compared to classical RHD solvers, we now have additional  
 equations on the number density of each specie $n_x$, which correspond to the advection, chemical creation and destruction of the $x$ chemical elements.  The dust grains, whether charged or not, are treated in exactly same way as the chemical species: They are advected with the gas velocities. We consider here neither a possibility of dust having a different velocity, nor the dust enrichment within the cloud \citep{bellan08}. 
On the right hand, the negative term is the sum over the destruction reactions, and positive term is the sum
over the reactions creating  $x$-species from species $p$ and $q$. $k_{rx}$ and $k_{pq}$ are the
corresponding reaction rates.
The cooling via molecular and atomic lines is present in {\ttfamily PDS} chemical code.
In the context of this paper,
we neglect the thermal feedback of chemistry on the gas to concentrate on which parameters are affecting 
the coolant's abundances.

\paragraph{The chemistry module}
We use the time-dependent chemistry {\ttfamily PDS} code, 
originally written for MHD shocks \citep{flower15,flower03,lesaff04,lesaffre05,lesaffre13}.
We solve the right side of the system of equations for chemical evolution of species $n_{x}$ (see last line in eqs.~\ref{chem_solver}) using either explicit or implicit integration in time. 
When the chemical time step is shorter than the hydrodynamical one, we use an implicit solver which consists in inverting the matrix 
of $N_\mathrm{species}^2$ in size for a hydro(-magnetic) time-step of $\Delta t_\mathrm{hydro}$. Should this not be the case, 
the chemical module switches to an explicit second-order Runge-Kutta method for time advancement.
In the implicit scheme, we use the DVODE solver\footnote{from the ODEPACK package downloadable at https://computation.llnl.gov/casc/odepack/}, without any optimisation of the matrix inversion procedure.

\paragraph{Methods in AMR RMHD code {\ttfamily RAMSES}} 
For the radiation transfer, we use the grey flux-limited diffusion (FLD) approximation 
described in detail in \citet{comm11a}, which combines the explicit second-order Godunov solver of {\ttfamily RAMSES}\rm\ for the hydrodynamical part, 
and an implicit scheme for radiative energy diffusion and coupling between matter and radiation 
terms. The implicit FLD solver uses  adaptive time stepping \citep[ATS, ][]{comm14}. 

 The MHD part of the code is based on the Constrained Transport (CT) scheme \citep{teys06}, 
using a 2D-Riemann solver to compute the electro-motive force at cell edges \citep{From06,teys06}.
In addition to the ideal MHD solver, ambipolar diffusion and Ohmic diffusion are also implemented as additional electro-motive forces in the
induction equation \citep{mass12}.  The aim of this paper is to estimate the relative importance of various magnetic diffusivity terms, depending on the fixed magnetic field strength, on the progress of the collapse and on the dust properties.

\paragraph{Current limitations} The {\ttfamily PDS} chemical code has no impact on the evolution of the gas in RHD part of {\ttfamily RAMSES}. 
The possible limitations are then three-folds. First, the total gas and dust density $\rho$ is not recomputed from the abundances obtained by the chemistry solver. 
In this paper, we consider FHSC before the dust melting occurs above 800~K. Second, as mentioned previously, we do not include heating and cooling due to atomic and molecular lines, which should not be dominant as long as radiative transfer is dominated by the dust. 
Last, we do not compute self-consistently the dust opacity from the dust composition given in output by the chemistry solver.
 For the latter, we instead use tabulated opacities from \citet{sem03}. 
All these limitations could be dealt by the two codes, but go far beyond the scope of the paper.

\section{Dense core collapse calculations: the setup}

\subsection{Initial conditions \label{sec:condinit}}

\begin{table}[t]
\vspace*{3mm}
\begin{tabular}{lllll}
\hline
Model  & $\sqrt{\langle{a^2\rangle}}$ & $n_\mathrm{dust}/n[{\rm H}]$ & $\langle{a^2\rangle}n_{\rm dust}/n[{\rm H}]$ \\
       &   [$\mu$m] &                                &  [$(\mu$m$)^2$] \\            
\hline
S1   & 0.1 & 3.1$\times{}10^{-12}$ & 3.1$\times{}10^{-14}$   \\ 
S2$_{\rm MRN}$   & 0.05 & 3.9$\times{}10^{-12}$ & 1.0$\times{}10^{-14}$   \\
S3$_{\rm MRN}$    & 0.017 & 5.2$\times{}10^{-11}$ & 1.5$\times{}10^{-14}$   \\
S4    & 1.00 & 3.1$\times{}10^{-15}$   & 3.1$\times{}10^{-15}$ \\
S5   & 0.02 & 3.9$\times{}10^{-10}$  & 1.4$\times{}10^{-13}$   \\
\hline
\end{tabular}
\caption{ \label{tab:collap}  List of 3D chemo-dynamical calculations of collapse. 
   The cloud is assumed to be always of one solar mass. S1 is a fiducial model. 
   S[2,3]$_{\rm MRN}$ are the models with MRN dust size distribution, whereas
    models S[1,4,5] have fixed grain size. $n_{\rm dust}{\langle{a^2\rangle}}$ represents mean dust cross-section. 
    }
\end{table}

\paragraph{Parameters of the core}
We choose a spherically-symmetric collapse
configuration, i.e. we neglect rotation, magnetic fields and turbulence. 
As we have mentioned before, the aim of this paper is to demonstrate the impact of dust properties on the magnetic diffusivities during the collapse. We make use of spherical symmetry and present easy-to-read 1D  interpretations of the results. 
The initial core mass is fixed to 1 M$_\odot$ and the temperature of both gas and dust is uniform 
and equals 10 K. 
Note that in our model, we assume that the dust and gas are perfectly coupled thermally. 
The adiabatic index  is $\gamma=5/3$ and the mean molecular weight is $\mu_\mathrm{gas}=2.375$. 
In this paper we use both $n_{\rm gas}=\rho_{\rm gas}/(\mu_{\rm gas}m_p)$ and $n[{\rm H}]=n_{\rm gas}/\mu_{\rm gas}$, 
where $m_{\rm p}$ is the mass of proton. 
We represent the relative abundance of $x$ chemical specie as $n[x]/n[{\rm H}]$.

In all our models, the initial density profile is Bonnor-Ebert like $n= n_\mathrm{c}/(1 + (r/r_\mathrm{c}))^{-2}$,
where the maximum density in the center $n_\mathrm{c}$ is 10 times larger than the density at the core's border. 
The total length of the simulation box is four times larger than the core initial radius $r_0$.  
The central density is $n_\mathrm{c}=4.4 \times 10^{5}$ cm$^{-3}$ (or, $1.71\times10^{-18 }$ g cm$^{-3}$), and the core radius is $r_0=0.022$ pc. 
The relation between thermal and gravitational energies is $E_{\rm th}/E_{\rm grav}=0.447$.

Assuming the same core initial properties as model S1, we vary the dust properties for the models S2$_{\rm MRN}$ to S5. 

\paragraph{Generating initial chemical abundances \label{init-cond}}
We adopt a uniform high visual extinction $A_\mathrm{v}=30$ and a uniform, cosmic ray ionisation rate of $1.3\times{}10^{-17}$ s$^{-1}$.
 In order to generate the initial
chemical abundances, we start from elemental abundances (see Paper I)
and let the 
chemistry evolve for $6\times{}10^5$ years in the static Bonner-Ebert sphere configuration  \citep[e.g.][]{hin13}. As mentioned previously, the assumptions about dust size distribution differs depending on the model (see Table~\ref{tab:collap}).

\paragraph{Introducing zero-time}
As already introduced in Paper I,  we define here as a zero-time $t_0$ the moment of the FHSC formation, i.e. when central density reaches $10^{13}$ cm$^{-3}$ ($T\sim 210$ K). The zero-time can appear different for the collapse simulations of various spatial expend (see Table~\ref{tab:collap} in Paper I). In the present paper, the time pace of the collapse in identical for all models, so that FHSC is formed at $t_0=57.8$ kyrs after the beginning of the simulation.

\subsection{Dust properties  \label{sec:dust} }

We compare five models with different dust size properties: models S1,
S2$_{\rm MRN}$, S3$_{\rm MRN}$, S4, and S5 (Table~\ref{tab:collap}). S2$_{\rm MRN}$ and S3$_{\rm MRN}$ models make use
of the MRN distribution for the grain sizes in the chemistry module. 
Models S4 and S5 have a fixed grain size (as in fiducial model S1) and serve as boundary cases, where we assume very large and tiny dust.  Note that all the five models use identical initial conditions and dust-to-gas ratio $f_\mathrm{dg}=0.01$, except thus for the dust size properties. 
We refer the reader to Appendix~\ref{sec:ADust} for a more complete description of the treatment of dust size. In the following, we briefly summarise the differences between these five models:
\begin{description}
\bf \item[S1]\rm: Fiducial case has single-sized dust with $a_0=0.1\mu$m, 
dust-to-gas ratio, and dust density $n_\mathrm{dust}=3.09\times10^{-12} n_\mathrm{H_2} $;
\item[\bf S2$_{\rm MRN}$\rm]: 
  We use the classical MRN dust grain size distribution $n_{\rm dust}(a)\sim{} a^{-3.5}$ \citep{mathis77},
  scaled to match the results of coreshine modelling \citep{ste10}. The mean dust radius is
  $\sqrt{\langle{a^2\rangle}}=0.05$ $\mu$m, what corresponds to $a_\mathrm{min}=2.6\times{} 10^{-6}$ cm 
  and $a_\mathrm{max}=5\times{}10^{-5}$ cm. 
  Fixing $f_\mathrm{dg}=0.01$, the dust number density is calculated to be
   $n_\mathrm{dust}=3.9\times{}10^{-12}n_\mathrm{H_2} $ 
(see Appendix~\ref{sec:ADust} for details);
\item[\bf S3$_{\rm MRN}$\rm]:  
  Same as S2$_{\rm MRN}$, but  
   we adopt  $a_{\rm min}=10^{-6}$~cm and  $a_{\rm max}=3\times{}10^{-5}$~cm \citep{flower03}. 
   The mean dust radius is $\sqrt{\langle{a^2\rangle}}=0.017\mu$m, 
   and $n_{\rm dust}=5.24\times{}10^{-11}n_\mathrm{H_2}$. 
This size range resembles the fit for mixed composition 'large' grains, made of bulk carbonaceous material, 
close to the range considered in \citep[][PAH grains are not considered]{mathis77};
\item[\bf S4\rm]: Same as S1, i.e. single-sized dust, with $a_0=1\mu$m, and $n_{\rm dust}=3.09 \times{}10^{-15}n_\mathrm{H_2}$; 
\item[\bf S5\rm]: Same as S1, i.e. single-sized dust, with $a_0=0.02\mu$m, and $n_{\rm dust}=3.87 \times{}10^{-10}n_\mathrm{H_2}$.
\end{description}

The radial profiles of selected chemical species for models S1, S2$_{\rm MRN}$, and S3$_{\rm MRN}$ at time $t_0$ are demonstrated in  Paper I. The conclusion was that the dust properties, such as size distribution and total number density, have a significant impact on the resulting abundances of all species, especially ions and other charged species.

\subsection{Magnetic diffusivities}
As we mention in the introduction, the chemistry of collapsing clouds has direct consequences on the dynamics via non-ideal MHD effects. We calculate the Ohmic, Hall and ambipolar diffusivities ($\eta_\mathrm{O}$, $\eta_\mathrm{H}$, and $\eta_\mathrm{A}$, respectively) from the number densities of charged
species obtained with our chemistry module \citep{war07}
\begin{equation}\label{eq:eta}
  \eta_\mathrm{O}=\frac{c^2}{4\pi\sigma_\mathrm{O}},
  \ \eta_\mathrm{H}=\frac{c^2\sigma_\mathrm{H}}{4\pi\sigma_\perp^2}\ \ {\rm and}\
  \ \eta_\mathrm{A}=\frac{c^2\sigma_\mathrm{P}}{4\pi\sigma_\perp^2}-\eta_\mathrm{O},
\end{equation}
with $\sigma_\perp=\sqrt{(\sigma_\mathrm{H}^2+\sigma_\mathrm{P}^2)}$,
where $\sigma_\mathrm{O}$, $\sigma_\mathrm{H}$, $\sigma_\mathrm{P}$ are the Ohmic, Hall and
Pedersen conductivities \citep{cow76,war99}
\begin{equation}\label{eq:sigmaO}
  \sigma_\mathrm{O}=\frac{ec}{B}\sum_xn_x|q_x|b_x,
\end{equation}
\begin{equation}\label{eq:sigmaH}
  \sigma_\mathrm{H}=-\frac{ec}{B}\sum_x\frac{n_x q_xb_x^2}{1+b_x^2},
\end{equation}
\begin{equation}\label{eq:sigmaP}
  \sigma_\mathrm{P}=\frac{ec}{B}\sum_x\frac{n_x|q_x|b_x}{1+b_x^2},
\end{equation}
where $B$ is the magnetic field, $n_x$ and $q_x$ are number density and charge of $x$ specie. 
The parameter $b_x=\tau_x\omega_x$
 describes how well the charged particles couple to the magnetic
field. It is calculated by putting together the cyclotron frequency $\omega_x$ and the
collisional damping time $\tau_x$.
A charged specie $x$ has cyclotron frequency
\begin{equation}
  \omega_x=|q_x|eB/(m_xc),
\end{equation}
where $q_xe$ is the charge and $m_x$ the mass.  The damping time
of the charged specie's motion relative to the neutrals is
\begin{equation}
  \tau_x=\frac{m_n+m_x}{m_n}\frac{1}{n_n\langle{\sigma v\rangle}_{xn}},
\end{equation}
where $\langle\sigma v\rangle_{xn}$ is the momentum-transfer rate
coefficient for colliding with neutrals, averaged over the Maxwellian
velocity distribution.  We adopt $\langle\sigma v\rangle_{\mathrm{en}} =
10^{-15}\sqrt{128 k_\mathrm{B}T/(9\pi m_\mathrm{e})}$ for the electrons, and
$\langle\sigma v\rangle_{\mathrm{in}} = 1.9\times 10^{-9}$ for the ions
\citep{dra83}.  The corresponding rate coefficient for the dust grains
follows eq.~21 in \citet{war99}, where the collision speed is the
neutrals' thermal speed since the grains' drift through the gas is
subsonic.  For electrons, typically $b_\mathrm{e}\gg 1$.  The quality of coupling
to the magnetic fields drops with the mass of the charged specie \citep[see for details][]{dzyur13}. 

Since our simulations are done without magnetic fields, 
we assume that the magnetic pressure scales with gas pressure as $5B^2/(4\pi)=nk_\mathrm{B}T$ (plasma beta $\beta=P_{\rm gas}/P_{\rm mag}=5$).
We then calculate the cyclotronic frequencies and friction times for each charged specie, ions, dust and electrons. 
Those values are used to determine resistivities (eqs.~\ref{eq:sigmaO},~\ref{eq:sigmaH},~\ref{eq:sigmaP}). 
In the next section, we show the resulting ambipolar, Hall and Ohmic resistivities, 
along with radial distribution of charge carriers. We focus only on the comparison of the models with various dust sizes. Changing  the initial core radius, i.e. the tempo of collapse, is not affecting the ionisation equilibrium for the chosen parameter range. 
%


\subsection{Fiducial model: evolution of magnetic diffusivities during the collapse \label{sec:fidu}}

\begin{figure*}
\begin{center}
\includegraphics[width=7.5in]{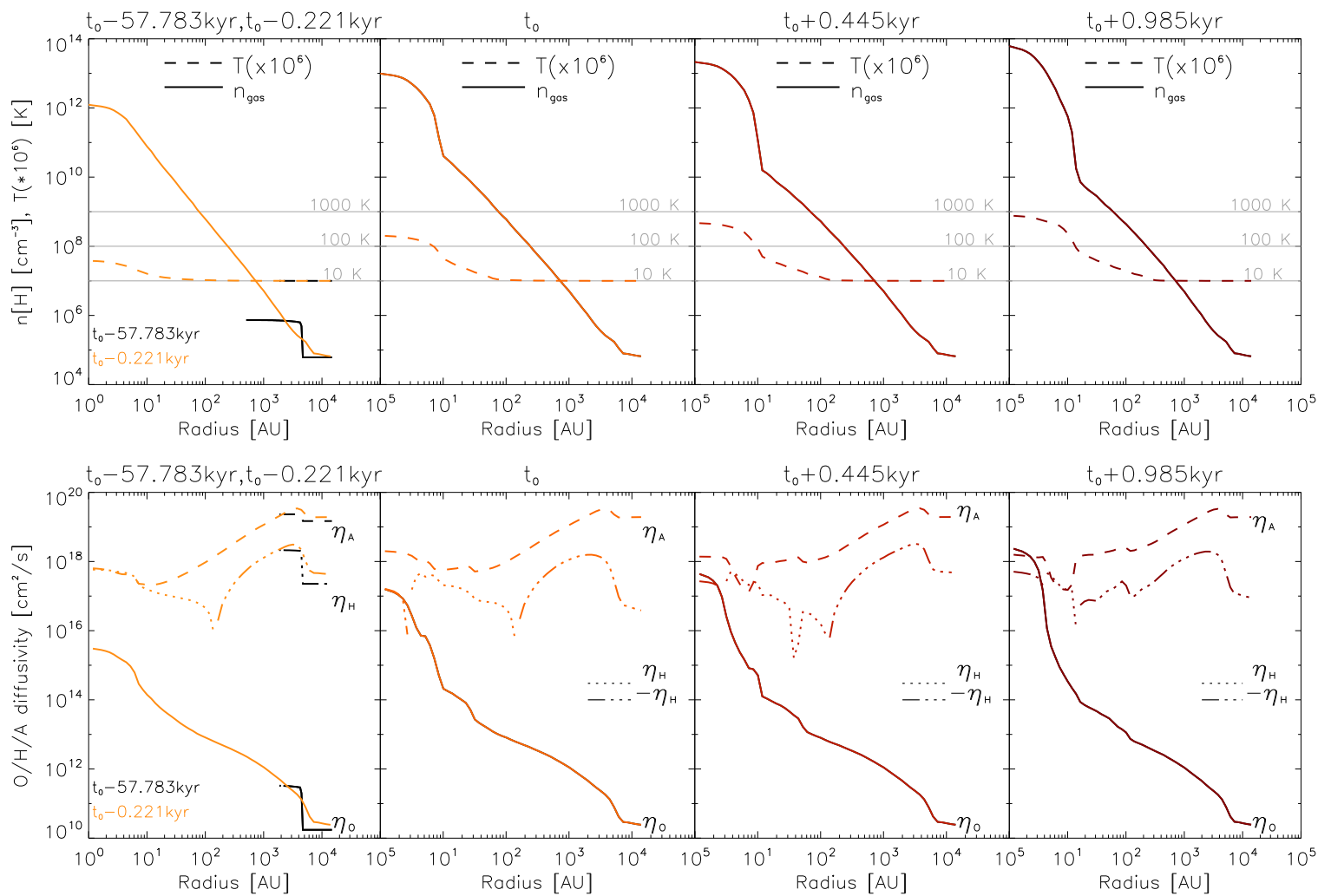}  
  \caption{ Time evolution of gas density, temperature and magnetic diffusivities in the fiducial collapse model, shown for five snap-shots. Plots on the left show both initial condition (in black) and the data at $t_0 -0.221 $kyrs which correspond to the beginning of the adiabatic collapse phase. Plots on the right show the last output, when we stop the simulation. 
}
\label{col1}
\end{center}
\end{figure*}
Figure~\ref{col1}~ shows the time evolution of the gas density, temperature and corresponding magnetic diffusivities
for five snap-shots, starting from initial condition and ending with the final output. The fiducial run is stopped when temperature reaches above 800 K. This is a temperature when a chemo-erosion of C from the dust cores has to be included. It can be expected that chemo-erosion will greatly affect the chemical abundances, because of large amount of C released in the gas phase. 

We remind that the details on the chemical abundances can be found in Paper I. Here we analyse the main trends in the radial behaviour of non-ideal MHD terms.
The Ohmic dissipation grows from $10^{12}$ to few of $10^{18} \rm cm^2/s$ through the simulation. 
The ambipolar diffusion is shown to dominate other magnetic diffusivities for the most time snap-shots of the fiducial simulation, with two exceptions.  At $t_0-0.221$ we observe that a negative Hall effect is at least as strong as the ambipolar diffusion within 10 AU (i.e, within FHSC radius). Second, at the final output we observe that Ohmic dissipation begins to dominate in the innermost 2-3 AU.
The sign reversal of the Hall effect will be discussed and explained in the following sections.

\section{Effect of dust properties \label{sec:dustdiff}}

\begin{figure*}
\begin{center}
\includegraphics[width=7.in]{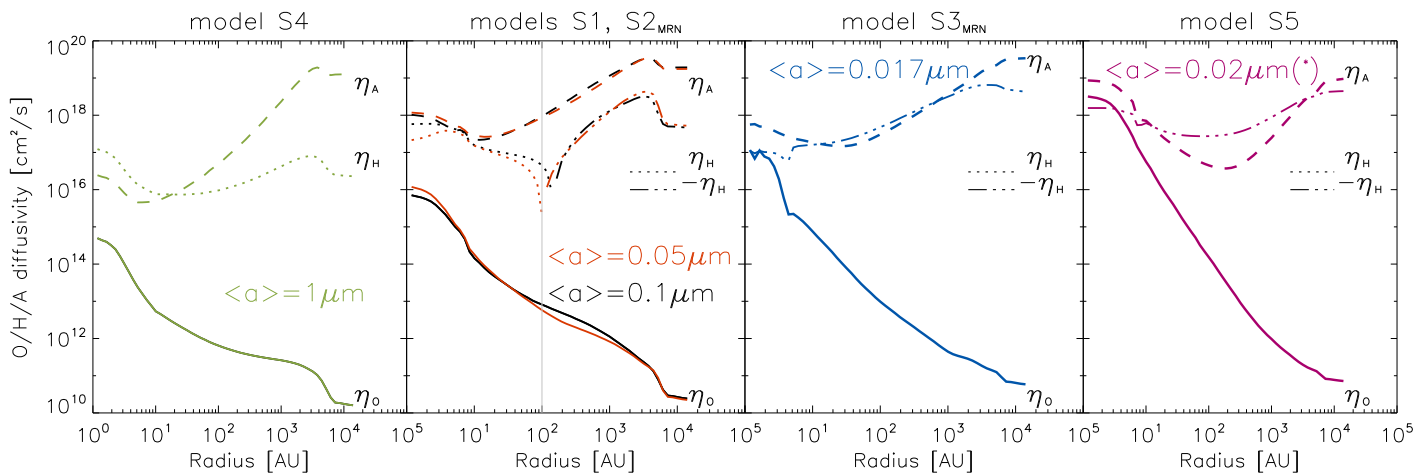}   
  \caption{ Magnetic diffusivities for different dust sizes: Ohmic diffusivity (solid line), 
          Hall diffusivity (dotted for positive and dash-dotted for negative) 
           and ambipolar diffusivity (dashed line) versus radius. 
           The plots are made  at time  $t_0-0.22$ kyrs, when central temperature is $T_c=55$~K 
           and central density 1.2$\times{10^{12}}$~cm$^{-3}$. 
           Asterix is reminding that the mean dust cross-section is minimal in model S5, and the mean dust radius is
           not an unique characteristic of dust properties. 
 }
\label{etas}
\end{center}
\end{figure*}

In this section, we analyse how the dust properties affect the charged species and thus the magnetic dissipation.
Ohmic, Hall and ambipolar diffusivities are plotted in fig.~\ref{etas}, where we compare the outcome of simulations at the moment of FHSC formation.
There, the models are arranged according to the mean dust size: Model S4 on the left with $\langle{a\rangle}=1\mu$m, followed by fiducial model S1 and S2$_{\rm MRN}$ with 0.1$\mu$m and  0.05$\mu$m, whereas models S3$_{\rm MRN}$ and S5 are on the right with mean dust sizes   0.017$\mu$m and  0.02$\mu$m. 
Note, that even mean dust size appear to be slightly larger in S5 model, 
its mean dust cross-section is the largest and closest to the value used in \citet{mar16} 
(see also fig.~\ref{meanopacity}).

Model S4 with mean dust size of 1$\mu$m has the lowest dust number density. The ambipolar diffusion is nevertheless almost as strong as in the fiducial model in the envelope for $r>10^4$AU, where it exceeds $10^{19} \rm cm^2 s^{-1}$. Towards the center of the cloud, the magnetic diffusivities are getting weaker when compared to the fiducial model: Ohmic resistivity drops to the few of $10^{14}\rm cm^2s^{-1}$, and ambipolar diffusion is about $10^{15}\rm cm^2s^{-1}$ within FHSC. Interesting to note that Hall diffusivity is dominating within $10$AU, also it is not reaching above $10^{17}\rm cm^2s^{-1}$ in the center of the cloud. We can summarise here that in case when dust is evolved and can be represented with averaged dust size of $1\mu$m, the gas is close to ideal MHD state and the rotationally-supported disk (RSD) cannot be formed. 

Model S2 appears to be very close to the fiducial S1 model. The ambipolar diffusion dominates over the whole radial extent, but with its maximum value of $10^{18}\rm cm^2s^{-1}$ it is probably not high enough to enable the formation of RSD around FHSC.  Interesting to note that negative Hall effect become as strong as the ambipolar diffusion between 4 and 8 AU. At the moment of FHSC formation, the Ohmic resistivity is a few orders of magnitude lower in the center of the cloud. 

Models S3$_{\rm MRN}$ and S5 appear to have close values of the mean dust size, but model S5 has larger dust number density and thus it is placed on the right in fig.~\ref{etas}.
Model S5 has the largest mean grain cross-section per hydrogen nucleus, which leads to 
the largest magnetic diffusivities:
Ohmic dissipation exceeds $10^{18}\rm cm^2 s^{-1}$ in the very center of the cloud (see fig.~\ref{etas}); Hall effect is negative over the whole radial extend and dominates over ambipolar diffusion between $10\rm AU$ and $10^3$AU. 
Only in S5 model, the ambipolar diffusion is reaching $10^{19}\rm cm^2 s^{-1}$ in the cloud center.

From comparing the magnetic diffusivity profiles in fig.~\ref{etas}, 
we can conclude that the chances to form RSD are increasing with the increasing mean dust opacity. This means that the small dust have to remain numerous regardless which size the largest dust grains may achieve. 
We point out that the Hall effect is switched to be negative for the small dust sizes and shall be included in the non-ideal RMHD simulations of the collapsing cloud.
In order to understand the impact of the dust properties better,
we can ask the following questions:
(1) What causes the reversal in the Hall effect? (2) Why ambipolar diffusion has always a rise within FHSC, and why it has a minimum in between $10^2$ and $10^3$AU in case of small dust? (3) Can we make a simplification in the calculation of the ambipolar diffusion without loss of accuracy, and use a constant ion mass like in \citet{oku09a} instead of solving a chemical network?
We address those questions below.
\begin{figure*}
\begin{center}
\includegraphics[width=7.in]{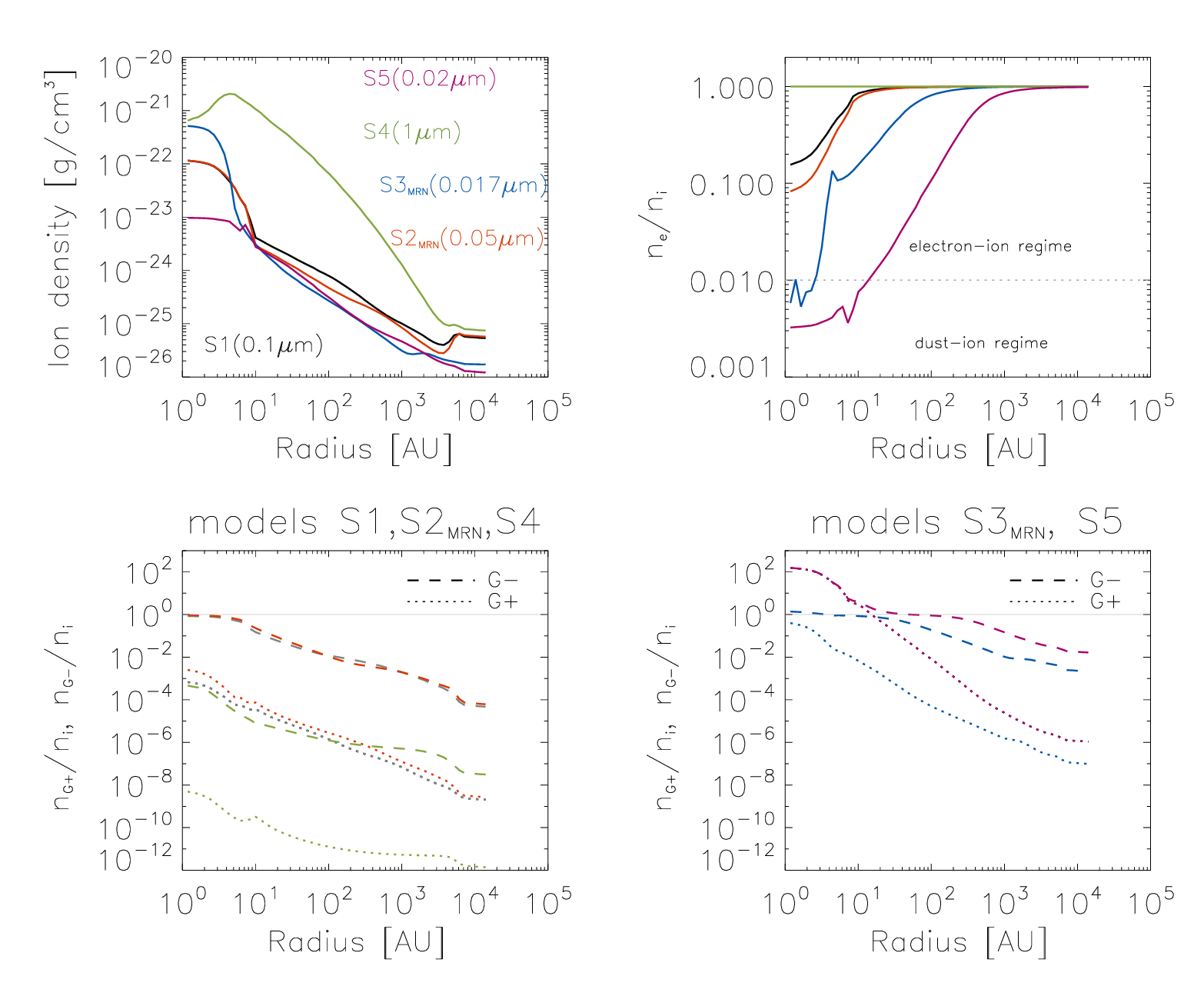} 
  \caption{ Various properties of charged species and magnetic diffusivities radial profiles, 
            shown for models S1 (black), S2$_{\rm MRN}$ (red), S3$_{\rm MRN}$ (blue), S4 (green) 
            and S5 (lilac). The colors refer to the same models in all following figures.
            Plots shows, from left to right: The density of ion versus radius, 
           the relation of electron density to the ion density,
           and at the bottom the relative abundances of charged dust to the ion density. }
\label{col5}
\end{center}
\end{figure*}

\subsection{On the sign reversal of the Hall effect}
Due to the total charge neutrality condition $q_en_e+q_in_i +q_{\rm G-}n_{\rm G-}+q_{\rm G+} n_{\rm G+}=0$ with $q_{x}$ is the charge of $x$-specie, 
there are only three 
independent parameters which can be defined as $n_e/n_i$, $n_{\rm G-}/n_i$, and $n_{\rm G+}/n_i$ \citep[see also sec.2 in][]{xubai15}.
In fig.~\ref{col5} we show  the dimensionless parameters  $n_e/n_i$, $n_{\rm G-}/n_i$,  and $n_{\rm G+}/n_i$.
These parameters show how the smaller dust size leads to larger inequality  $n_{i}> n_{e}$ and the transition to the dust-ion regime and even dust-dust regimes, i.e. when main charge carriers are ions and $\rm G-$ (model S3$_{\rm MRN}$), or $\rm G+$ and  $\rm G-$ (model S5).

The Hall effect term, given in eq.~\ref{eq:eta}, becomes negative when the Hall conductivity is negative, $\sigma_\mathrm{H}<0$. 
In case of single-charged dust and local charge neutrality,  eq.~\ref{eq:sigmaH} can be rewritten as 
\begin{equation}
 \sigma_\mathrm{H}=\frac{ec}{B}\left(-\frac{n_{\rm e}}{1+b_{\rm e}^2}+\frac{n_{\rm i}}{1+b_{\rm i}^2}+\frac{n_{\rm G+}-n_{\rm G-}}{1+b_{\rm G}^2}\right),
\end{equation}
where we reduce the dust coupling parameter to $b_{\rm G}$ because it is insensitive to the sign of the charge, i.e.  $b_{\rm G+}=b_{\rm G-}$.

\begin{figure*}
\begin{center}
\includegraphics[width=7.in]{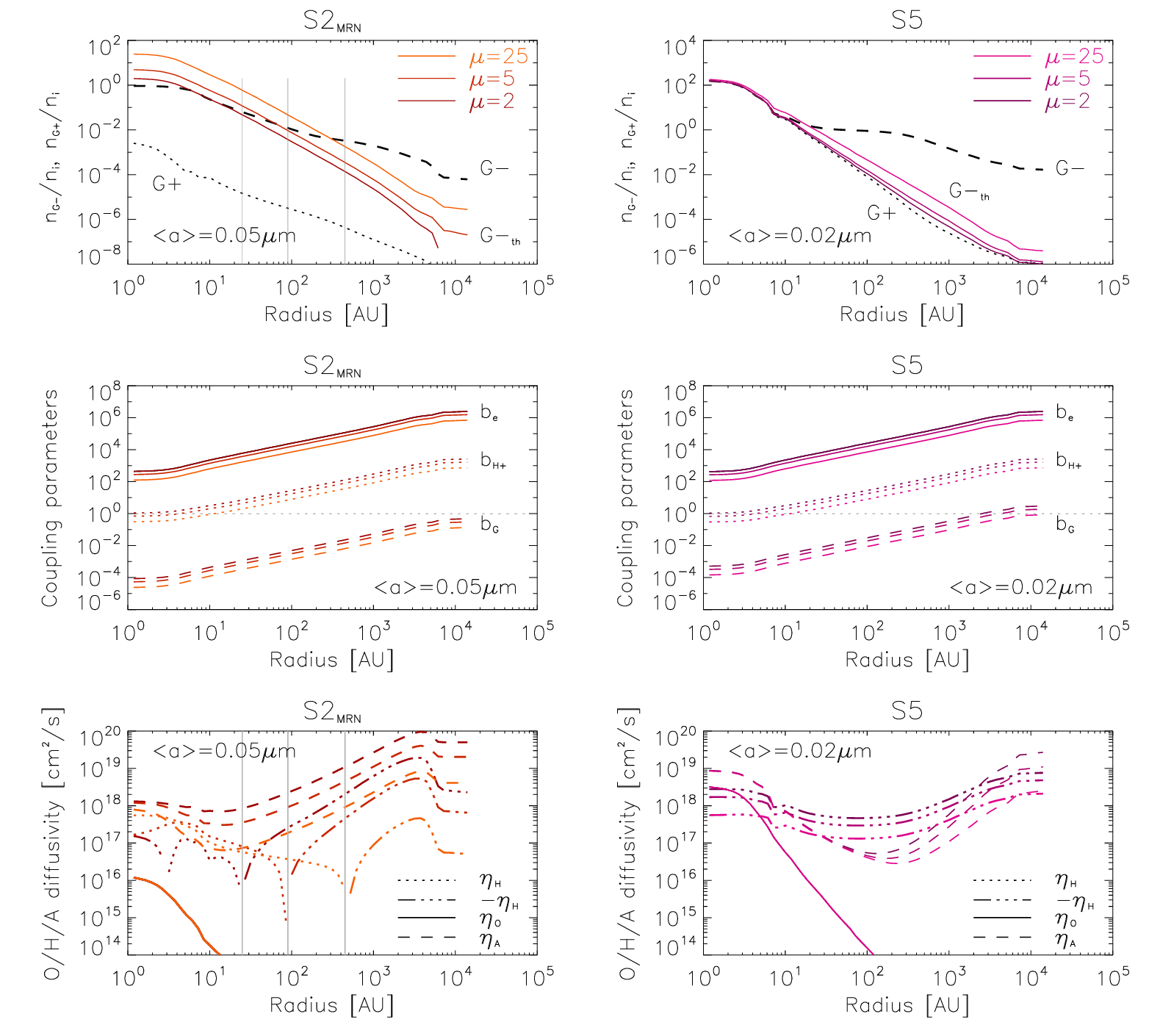} 
  \caption{ Top: Relative number densities of positively and negatively charged dust, $n_{\rm G+}/n_{\rm i}$ and $n_{\rm G-}/n_{\rm i}$, compared to the threshold value in eq.~\ref{eq:thres}.  Middle: Coupling parameter $b$ for electrons, $\rm H^+$ ions and dust (eqs.~\ref{eq:sigmaO}-~\ref{eq:sigmaP}). Bottom: magnetic diffusivities depending on the magnetic field strength, with colors showing $\mu=2,5,25$.
 }
\label{col6}
\end{center}
\end{figure*}

In our simulations, we observe that the coupling parameters $b_{\rm e}\gg{}1$ everywhere, and $b_{\rm i}\gg{}1$ and $b_{\rm G}\ll{}1$ for radii $10\rm AU <r <10^4 $AU (see also fig.~\ref{col6}). Thus, we simplify the equation for $\sigma_{\rm H}$ further and derive the threshold value for $n_{\rm G-}^{\rm th}$,
\begin{equation}
n_{\rm G-}^{\rm th}=\frac{n_{\rm i}}{b_{\rm i}^2}- \frac{n_{\rm e}}{b_{\rm e}^2}+n_{\rm G+} \cong{}\frac{n_{\rm i}}{b_{\rm i}^2} +n_{\rm G+}.
\label{eq:thres}
\end{equation}
When $n_{\rm G-}>n_{\rm G-}^{\rm th}$, the Hall effect changes sign to the negative which may lead to the interesting MHD effects during the collapse. {\it The Hall effect becomes negative, when the number of negative charge carries, weighted over coupling parameter $b_{x}^2$, is dominating  the number of $b_{x}^2$-weighted positive carriers}. This is likely to happen when both the negatively charged dust is abundant, and the magnetic fields are strong enough for the strong coupling of ions ($b_i>1$).

From last equations we learn that both the number of negatively charged dust and also the 
quality of coupling to the magnetic fields are important. 
It is now a natural question to ask what happen if we change the cloud magnetisation. 
We consider cases with $\beta_{\rm plasma}=4\pi nk_\mathrm{B}T/B^2=[2,5,25]$.
The various magnetisation is shown with colors in fig.~\ref{col6} for models S2$_{\rm MRN}$ and S5.
We observe that the radial position of the sign reversal in Hall effect is shifting from outside to inside, when the magnetic field is increased:  $450$AU for $\beta_{\rm plasma}=25$,  $90$AU for $\beta_{\rm plasma}=5$ (fiducial case) and  $25$AU for $\beta_{\rm plasma}=2$.
At the same time, the Hall effect gets stronger in its absolute value, getting closer to ambipolar dissipation in case of S2$_{\rm MRN}$, or overtaking in case of S5. 
As we see, the coupling parameter $b_{x} $ is changing for different magnetisation, but the common trend of $b_{\rm e}\gg{}1$ and $b_{\rm i}\gg{}1$ and $b_{\rm G}\ll{}1$ is kept for most of radial extend. We find that it is possible to have a reversal in sign of the Hall effect without necessary reaching $b_{\rm G}\cong{}1$ as in \citet{xubai15}. We remind here that \citet{xubai15} have been investigating the possibility of Hall effect reversal in the proto-planetary disks, where the coupling parameters for charged species can differ significantly compared to those in the collapsing cloud. 

\subsection{Ambipolar diffusion}

\begin{figure}
\begin{center}
\includegraphics[width=3.5in]{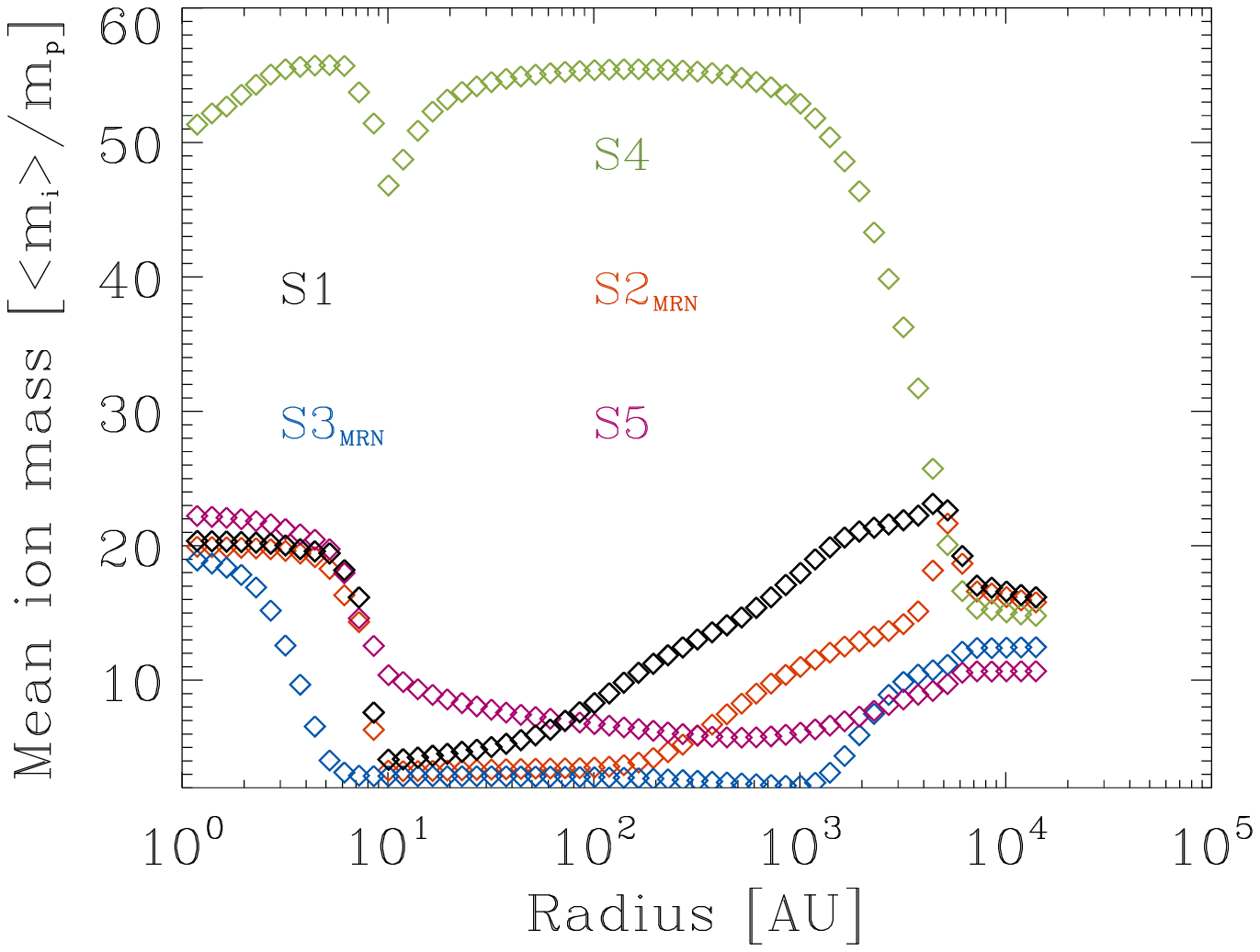}  
  \caption{ The snap-shot of mean ion mass $m_{\rm i}/m_{\rm p}=\rho_{\rm i}/(m_{\rm p}n_{\rm i})$ as a function of radius, 
       at time $t_0-0.22 \rm kyr $ (same time snap-shot as in figs.~\ref{col5}-\ref{col7}).
 }
\label{massI}
\end{center}
\end{figure}

The ambipolar diffusion is often implemented in form of $\eta_{\rm A}=B^2/(\gamma \rho \rho_{\mathrm i})$ in various numerical codes. From this formula, one can see that the ion density is expected to be mainly responsible for the ambipolar diffusion. There are 18 ions included in our network (see Paper I for details). Most massive are $\rm Fe^+$ ($55,85m_{\rm p}$) and $\rm HCO^+$ ($29m_{\rm p}$). In the envelope, $\rm CH_5^+$ is often the most abundant ion, accompanied by $\rm C^+$. From fig.~\ref{massI} we can see how the mean ion mass $<m_{\rm i}>=\rho_{\rm i}/n_{\rm i}$ is distributed along the radius. For fiducial case S1, the mean ion mass is about $17m_{\rm p}$ at radii $r>10^4$AU, 
peaking to $24m_{\rm p}$ at $r\simeq 5\cdot{}10^3$AU where the density jump from the initial condition was situated.
Then the mean ion mass is dropping to $3m_{\rm p}$ when going deeper into the cloud, until the border of the FHSC is reached. There, the temperature is reaching $55$K for the snap-shot we choose, so the thermal adsorption of most ices is happening and the mean ion mass reaches again $24m_{\rm p}$. There is no clear correlation between mean ion size and the mean dust size, but the trend is similar for all simulations but S4.

\begin{figure*}
\begin{center}
\includegraphics[width=7.in]{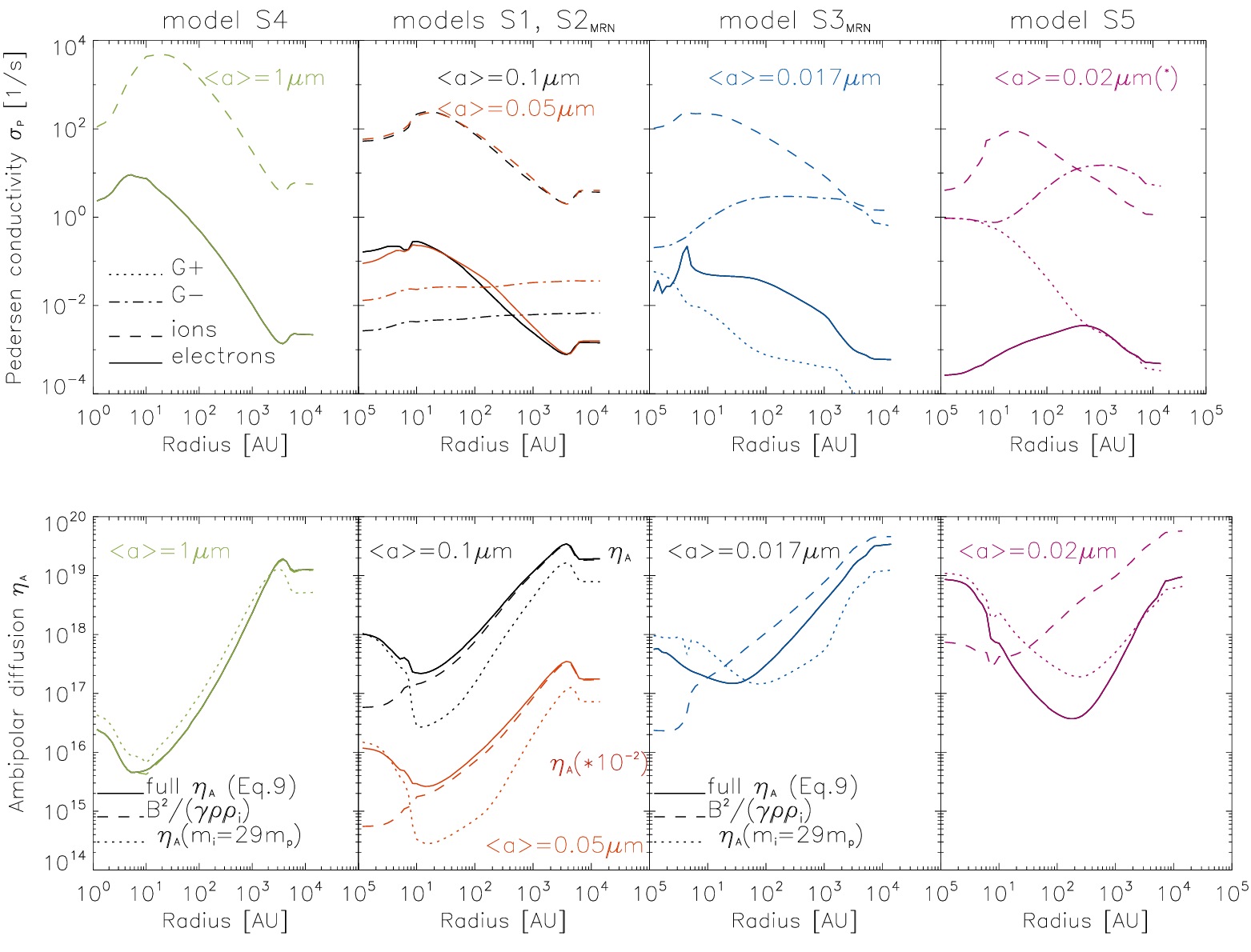}  
  \caption{ Top: Contribution to Pedersen conductivity for each type of charge carrier. 
      Bottom: ambipolar diffusion $\eta_{\rm A}$ according to eq.~\ref{eq:eta} shown with solid line, 
      $\eta_{\rm A}$ under the assumption of overall constant ions mass $m_{\rm i}=29m_{\rm p}$ shown with dotted line, and 
      a frequently-used $\eta_{\rm A}=B^2/(\gamma \rho \rho_{\rm i})$ is shown with dashed line. 
      Asterix in $\langle{a\rangle}=0.02\mu\rm  m(*)$ is to remind that mean dust radius is not a unique 
    parameter to characterise the properties of dust, and that model S5 has maximal mean dust cross-section.  
 }
\label{col7}
\end{center}
\end{figure*}

We plot the contribution to Pedersen conductivity for each type of charge carrier in fig.~\ref{col7}, top. Bottom row in fig. ~\ref{col7} shows the ambipolar diffusion, calculated in three different ways. Solid line shows ambipolar diffusion $\eta_{\rm A}$ according to eq.~\ref{eq:eta}, using the ion, electron and charged dust abundances from the chemistry outputs. With dotted line we show the value of $\eta_{\rm A}$ when we assume that the number density $n_i$ is same as in chemical output, but the ion mass is constant everywhere (i.e. $m_{\rm i}=29\mathrm{m_p}$ to represent  $\rm HCO^+$) 
 and not a function of radius as shown in fig.~\ref{massI}. 
With dashed line we plot  $\eta_{\rm A}=B^2/(\gamma \rho \rho_{\mathrm i})$ in fig.~\ref{col7}(bottom), which would be a correct representation of ambipolar diffusion in the absence of dust, when $n_{\rm e}=n_{\rm i}$.

In model S4, the contribution of dust to Pedersen conductivity is negligible and $n_{\rm i}=n_{\rm e}$ everywhere (see also fig.~\ref{col5}). Thus, the dashed and solid lines are coinciding there. The assumption of a constant ion mass is not too bad either, only it should be iron with $m_{\rm i}=55.8m_{\rm p} $ instead of $\rm HCO^+$ to match the curve more accurately.  

For other models, the condition $n_{\rm i}=n_{\rm e}$ is not fulfilled and the impact of dust is therefore not negligible. The departure of dashed lines ($\eta_{\rm A}=B^2/(\gamma \rho \rho_{\rm i})$) from solid lines (showing $\eta_{\rm A}$ according to eq.~\ref{eq:eta}) start exactly at the radial location where also $n_{\rm i} \ne n_{\rm e}$ (see fig.~\ref{col5}).

We observe that ambipolar diffusion increases for smaller dust steadily from model S4 (left) to model S3$_{\rm MRN}$, but drops for model S5 where the dust is both most abundant and small. This drop occurs when the negatively charged dust is a stronger contributor to the Pedersen conductivity as the ions themself. By comparing with fig.~\ref{col5}, one can also see that S5 is the only model where both $\rm G+$ and $\rm G-$ are important charge carriers.

To summarise from the fig.~\ref{col7}, we need to solve a chemical network together with collapse in order to obtain an accurate radial distribution of ion masses. The attempts to simplify by using  $\eta_{\rm A}=B^2/(\gamma \rho \rho_{\rm i})$ formulation, or by using a simpler method to obtain the ionisation of the cloud  with fixed ion mass as in \citet{oku09a}, both may lead to one order in magnitude difference in the ambipolar diffusion coefficient.

\subsection{On the importance of magnetic dissipation mechanisms during the collapse}
For the adopted range in dust size and distribution, we summarise how the dust properties affect the magnetic diffusivities before FHSC formation:
(a) Ohmic dissipation increases strongly when the dust size is reduced, but it never dominates;
(b) Ambipolar diffusion always dominates in the envelope. It increases as well  when the dust size is reduced, but for dust sizes below 0.02$\mu$m ambipolar diffusion looses its importance against the Hall effect due to the charged dust contribution;
(c) Hall effect is gaining on importance when dust size is reduced, and for dust sizes between 0.1 to 0.05$\mu$m it shows the sign reversal. For mean dust sizes  smaller then $0.02\mu$m, Hall effect may become negative everywhere. 

Recently it has been shown that ambipolar diffusion alone can lead only to small disk sizes in the aligned rotator configuration  \citep{tsukamoto15,mas15}. The inclusion of Hall effect may change the situation dramatically: depending on alignment with magnetic field, large and more massive disks can be formed \citep{tsukamoto15b}. Those simulations are still in the need for verifications, in particular in the context of turbulent dense cores,  but these results combined with our findings stress once more how important is to consider the dust properties (size, charge, growth)  for the early stages of stellar formation.   

\subsection{Limitations}

There are several limitations in our dynamical chemical solver that can potentially affect the estimate of ionisation and thus the resulting magnetic diffusivities. We discuss below the three main limitations and defer their implementation in our model to future work.

First, we assume that the grain size distribution can be represented by a single mean size. \cite{kun09} show that the dust size distribution should be described by at least five bins to have a convergence of less than 1\% in the abundances. More recently, \cite{mar16} find that increasing the number of size bins from one to five can change the non-ideal MHD resistivities up to one order of magnitude, in particular within the density range where non-ideal effects are expected to play an important role. 

Second, we consider single-charged grains whereas they can hold multiple electric charges \citep{dra87,dzyur13}. Although multiple charged grains are scarce in dense cores \citep{ume80}, they play an important role in the ionisation budget (more multiple charged grains means less single charged grains). In the same line, \cite{mar16} show that charge transfer between grains has a first order effect for densities larger than $10^{10}-10^{11}$ cm$^{-3}$ and that it becomes even more important in the case of multiple charge. Currently, chemical models are limited by numerical difficulties because of the large dynamical range in the abundance of multiple charged grains, which is hard to handle in classical matrix inversion methods such as DVODE. Analytical models have been developed to treat multiple charges but they lack charge transfer between grains  \citep[e.g.][]{dra87}, or force the assumption of the constant ion mass \citep{oku09a}.

Last, we do not account for the variations of the CR ionisation rate deep inside the collapsing core. \cite{Padovani13,pad14} studied the propagation of CR along magnetic fields lines in collapsing cores, accounting for the effect of CR energy loss and magnetic mirroring. They show how CR ionisation rate is efficiently reduced in the central region because of the complex magnetic field lines configurations found in numerical models of protostellar collapse, making magnetic resistivities larger towards the cloud center.

\section{Conclusion}

We have performed 3D chemo-dynamical radiation-hydrodynamics simulations of 1 M$_\odot$ isolated dense core collapse.
Second, the computation of magnetic diffusivities can be done on the fly  in such chemo-dynamical collapse simulations. 

We use the chemical abundances resulting from our RHD simulations to study the impact of dust properties on the magnetic dissipation, and interpret the findings in the context of the previously performed collapse simulations resulting in the formation of the disks around the protostar.
%
%
We compare the diffusivities for the chosen range of dust sizes at the moment of FHSC formation, which we define as when cloud center has  $n_{\rm gas}=1.2\times10^{12} \rm cm^{-3}$ and $T=55$K. Our main findings are:
\begin{itemize}
\item{} We consider a range in the mean dust sizes, from 1$\mu$m to 0.017$\mu$m, assuming both single-sized dust and a dust-size distribution. The range of mean dust cross-section $\langle{a^2\rangle}n_{\rm dust}/n_\mathrm{H_2}$ is varied from $3.1\times10^{-15}$ to $1.4\times10^{-13} \mu\rm m^2$. Within this range of dust properties, the gas in the molecular cloud makes a transition from the ionisation state with $n_e=n_i$ in case of  1$\mu$m dust, up to the weakly ionised gas  where main charge carries are ions and $G-$, or only charged dust $G-$ and $G+$ in case of 0.02$\mu$m. 
\item{} For the chosen dust size range, the Ohmic diffusivity grows from $7\times10^{14}\rm cm^2 s^{-1} $ to $4\times10^{18}\rm cm^2s^{-1} $ at the cloud center.  Ohmic diffusivity is shown to be dominating magnetic dissipation first after the central gas temperature and density and temperature are reaching $n_{\rm gas}=10^{14} \rm cm^{-3}$ and $T=800$K (in the fiducial run).
For the disk formation it means that Ohmic dissipation alone would lead to the formation of the tiny disk within FHSC.
\item{} Hall effect is increasing of about one order in magnitude when dust size is reduced from 1$\mu$m to 0.017$\mu$m. We also find that for mean dust size of 0.1$\mu$m the sign of $\eta_\mathrm{H}$ becomes negative at $r>90$ AU, when assuming a moderate cloud magnetisation, $\beta=5$. The strength of magnetic field affects the radial location of the sign reversal, it is found at 25 AU for $\beta=2$  and at 450 AU for $\beta=25$. We find that the sign reversal occurs when both negatively charged dust is dominating and ions are well coupled to magnetic field, i.e. 
$n_{\rm G-}> n_{\rm i}/b_{\rm i}^2+n_{\rm G+}$.
\item{} For dust sizes 0.02$\mu$m and smaller, the Hall effect is negative overall in the cloud. We expect the effect becoming even more pronounces when dust-to-gas ratio is enhanced.  The reason is the strong contribution of charged dust to the conductivities, which may even overweight the contribution of ions. 
\item{} We find that ambipolar diffusion always dominates in the envelope in case of ISM-typical dust sizes and larger up to 1 $\mu$m. It is also reliably dominant for $r>10^4$AU, where it is equal or slightly above $10^{19}\rm cm^2 s^{-1}$. Towards the cloud center, its value increases from $10^{16}\rm cm^2 s^{-1}$ for mean dust size 1$\mu$m to $10^{17}\rm cm^2 s^{-1}$ for 0.05$\mu$m, but for smaller dust it is declining again and the Hall effect becomes dominant for the large radial range within the envelope.
\item{} We also find that the detailed knowledge of ion masses along the radius is important for obtaining the accurate value of ambipolar diffusion.
\end{itemize}
We conclude that the changing of the mean dust radius from 1$\mu$m to 0.02$\mu$m would
lead to at least 4 orders in magnitude change for the Ohmic resistivity in the cloud center, to the
 reversal and up to 2 orders in magnitude increase for the Hall effect, and
  up to 3 orders in magnitude increase for the ambipolar diffusion AD in the cloud center.
 \citet{mar16} results are complementary to our study, showing up to one order in magnitude  difference for the magnetic diffusivities caused by accounting for the multiple dust charging, the lower value of CR ionisation rate, the several size bins for dust. 

It is interesting to note, that we can separate two scenario of the cloud collapse. First one, in case when average dust size is equal or larger than ISM values, is clearly dominated by ambipolar diffusion and will result in relatively small disks forming according to \citet{henne16}. Second scenario would be possible when large amount of much smaller dust is leading to the Hall effect dominated collapse, where the results of \citet{tsukamoto15b} would apply including factor 10 larger disks when compared with ambipolar-diffusion-dominated scenario. We propose that the size of circumstellar disks formed in Class 0 objects should correlate with the size of dust and/or the enrichment in dust-to-gas ratio.

Comparing our findings with the previous works \citep{mar16,mas15,hincelin16} we can conclude that the including the dust growth model into the 3D RMHD simulations of the collapse is expected to be vital for our understanding of the non-ideal MHD processes during the star formation. 
%


\bibliographystyle{aa}
\bibliography{dzyurkevich2_2016}

\begin{appendix}

\section{Approximate treatment of the dust-size distribution \label{sec:ADust}}

In our chemical module, we want to take into account the effect of dust-size
distribution on the chemistry and to avoid high memory costs due to treatment of numerous
size bins. 
Here, we trace the number densities only of species G0, $\rm G-$ and $\rm G+$.
For simplicity, we neglect the multiple dust charges.
In case of fixed dust size, the density of dust grain is simply defined as
\begin{equation}
n_\mathrm{dust}=\frac{M_{\rm gas} f_{\rm dg} }{4/3 \pi a^3 \rho_{\rm solid} },
\label{app1}
\end{equation}
where $a$ is the grain radius and $\rho_{\rm solid} $ is the internal density of the solid materials. 
We adopt a MRN size distribution $dn/da \propto a^{-3.5}$ \citep{mathis77}.

Next, the dust radius can be fixed, like in models S1, S4 and S5, or one can choose a range of sizes:
the minimum grain radius is $a_{\rm min}=0.01$ $\mu$m and maximum 
radius $a_{\rm max}=0.3$ $\mu$m (models S2$_{\rm MRN}$ and S3$_{\rm MRN}$). 
For the chemistry, the rates of ion-dust or electron dust reactions are directly affected by 
number density of the dust. The total available
surface of dust grains affects the rates of the freeze-out of species on dust.
We can determine $n_\mathrm{dust}$ using Eq.~\ref{app1}
after calculating the effective cubic radius $\langle{a^3\rangle}$ from
\begin{equation}
\langle{a^3\rangle} = \frac{\int_{a_{\rm min}}^{a_{\rm max}}  a^3 a^{-3.5} da}{ \int_{\rm min}^{\rm max} a^{-3.5} da}= 5\frac{a_{\rm max}^{0.5}-a_{\rm min}^{0.5}}
           {a_{\rm min}^{-2.5}-a_{\rm max}^{-2.5}} .
\label{app2}
\end{equation}
Similarly, we can use the number-weighted squared radius $\langle{a^2\rangle}$ for reactions sensitive to dust surface, and define the dust mean opacity (see appendix~\ref{sec:crosssection}).
\begin{equation}
\langle{a^2\rangle}=2.5\frac{a_{\rm min}^{-0.5}-a_{\rm max}^{-0.5}}
            {a_{\rm min}^{-2.5}-a_{\rm max}^{-2.5}}.
\label{app3}
\end{equation}
In table~\ref{tab:collap}, the average grain radius is 
\begin{equation}
\langle{a\rangle}=2.5\frac{a_{\rm min}^{-1.5}-a_{\rm max}^{-1.5}}
            {a_{\rm min}^{-2.5}-a_{\rm max}^{-2.5}}/1.5.
\label{app4}
\end{equation}

\section{Mean dust opacity or cross-section \label{sec:crosssection}}

\begin{figure}
\begin{center}
\includegraphics[width=3.6in]{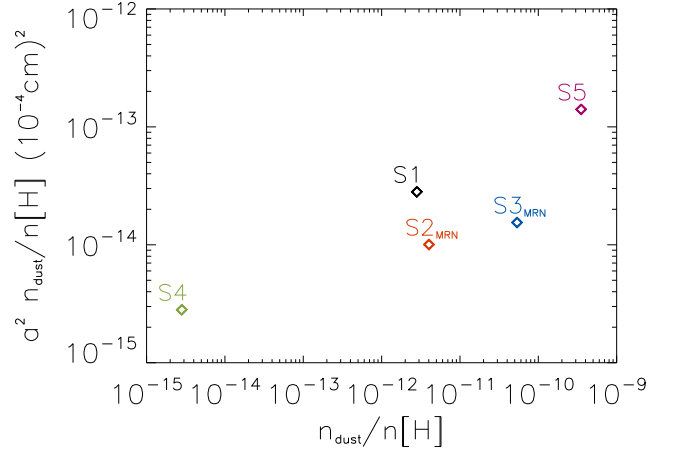}   
  \caption{ Mean grain cross-section per hydrogen nucleus as a function of dust density $n_\mathrm{dust}$ for models S1, S4, S5, S2$_{\rm MRN}$ and S3$_{\rm MRN}$. 
Model S2$_{\rm MRN}$ is taken closest to coreshine modelling.
 }
\label{meanopacity}
\end{center}
\end{figure}

We would like to emphasise here the importance of dust mean opacity $\langle{a^2\rangle}n_\mathrm{dust}$, as a value which also affects strongly the ionisation state of the gas.
The creation rate in case of the charge transfer or the recombination is
proportional to $n_\mathrm{dust}$. The creation rate for ice species is proportional to $\langle{a^2\rangle}n_\mathrm{dust}$. 
Whereas the dust density decreases steadily from model S1 to S3$_{\rm MRN}$, the parameter  
  $\langle{a^2\rangle}n_\mathrm{dust}$ is maximal for S1,
followed by S3$_{\rm MRN}$, with a minimum value in S2$_{\rm MRN}$ (see fig.~\ref{meanopacity}). 
Cases S4 and S5 represent low dust impact and high dust impact cases, correspondingly.

\end{appendix}

\end{document}